\documentclass[conference]{IEEEtran}
\IEEEoverridecommandlockouts

\usepackage{xcolor}
\usepackage{subfigure}
\usepackage{color}
\usepackage{graphicx}
\usepackage{balance}
\usepackage{amsmath}
\usepackage{amsfonts}
\usepackage{amssymb}
\usepackage{pifont}
\usepackage{times}
\usepackage{paralist}
\usepackage{textcomp}
\usepackage[hyphens]{url}
\usepackage[font={footnotesize}]{caption}
\usepackage{units}
\usepackage{enumitem}
\usepackage{amsmath,colortbl}
\usepackage{cite,comment}
\usepackage[nolist]{acronym}
\usepackage{soul}
\usepackage{array}
\usepackage{tabularx}
\usepackage{multirow}
\usepackage{lipsum}
\begin{acronym}[ACRONYM]
 \acro{3GPP}{the third generation partnership project}
\acro{5G}{fifth generation}
\acro{6G}{sixth generation}
\acro{AI}{artificial intelligence}
\acro{AoA}{angle-of-arrival}
\acro{AoD}{angle-of-departure}
\acro{BS}{base station}
\acro{CDF}{cumulative density function}
\acro{MIMO}{multi-input multi-output}
\acro{IoT}{internet of things}
\acro{RSSI}{received signal strength indication}
\acro{CSI}{channel state information}
\acro{SV}{Saleh-Valenzuela}
\acro{ULA}{uniform linear array}
\acro{BS}{base stations}
\acro{CNN}{convolution neural network}
\acro{MFM}{Matlab for MIMO}
\acro{CDF}{cumulative distribution function}
\acro{RTI}{radio tomographic imaging}
\acro{ISAC}{integrated sensing and communication}
\acro{CPI}{coherent processing interval}
\end{acronym}

           \newcommand{\mbf}[1]{\mathbf{#1}}          
\newcommand{\mb}[1]{\mbox{#1}}
\newcommand{\SUB}[2]{#1_{\scriptsize \mbox{#2}}}
\newcommand{\MS}[1]{\scriptsize \mbox{#1}}
\newcommand{\suchthat}{\, \mid \,} 
\newcommand\norm[1]{\left\lVert#1\right\rVert}

\begin{document}
\bstctlcite{IEEEexample:BSTcontrol}

\title{Multistatic Sensing of Passive Targets Using 6G Cellular Infrastructure}
\date{\today}
\author{Vijaya~Yajnanarayana\IEEEauthorrefmark{1}, Henk Wymeersch\IEEEauthorrefmark{2}\\
\IEEEauthorrefmark{1}Ericsson Research, India
\IEEEauthorrefmark{2}Chalmers University of Technology, Sweden\\
Email: vijaya.yajnanarayana@ericsson.com, henkw@chalmers.se
}
\maketitle

\begin{abstract}
  Sensing using cellular infrastructure may be one of the defining feature of sixth generation (6G) wireless systems. 6G communication channels  operating at higher frequency bands (upper mmWave bands) are better modeled using  clustered geometric channel models. In this paper, we propose methods for  detection of passive targets and estimating their position using communication deployment without any assistance from the target. A novel AI architecture called CsiSenseNet is developed for this purpose. We analyze  resolution, coverage and position uncertainty for practical indoor deployments. Using the proposed method, we show that human sized target can be sensed with high accuracy and sub-meter positioning errors in a practical indoor deployment scenario. 

\end{abstract}

\section{Introduction}
\label{sec:intro}

The \ac{6G} wireless systems will continue to evolve towards higher frequency bands and wider bandwidths \cite{yang20196g}. Typical 6G deployment will be spread over low, mid and higher frequency bands to enhance coverage and capacity \cite{rikkinen2020thz}. The increase  in operating frequency could result  in communication bands operating closer to traditional radar bands. We see this trend already in \ac{5G} mmWave communication bands merging with  K band ($18~\mb{GHz}-26.5~\mb{GHz}$) and Ka band ($26.5~\mb{GHz}-40~\mb{GHz}$) and this trend will continue in \ac{6G}. High frequency operation of 6G enables  transceivers to employ massive antenna arrays. This coupled with wider bandwidth can aid in high resolution sensing solutions with fine range, Doppler and angular resolutions \cite{wymeersch-2021-integ-commun, behravan-2022-introd}.

\begin{figure}
  \centering
        {\includegraphics[width = 0.9\columnwidth]{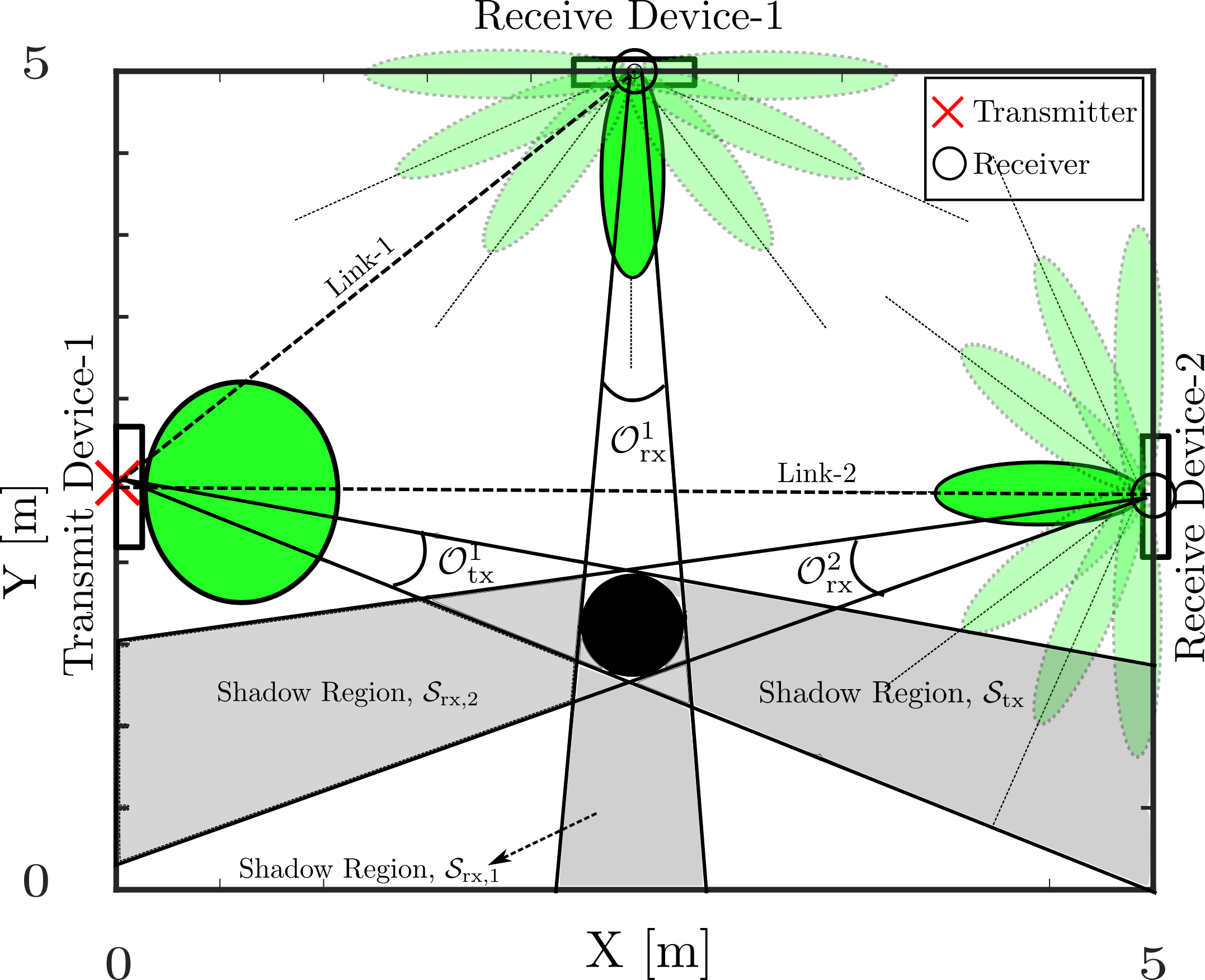}}
        \caption{Sensing of a passive object in an indoor wireless deployment. The passive object creates shadow regions which in turn creates perturbation in the communication link which can be exploited towards sensing. }
        \label{fig:occlusion} \vspace{-3mm}
    \end{figure}

As visualized in Fig.~\ref{fig:occlusion}, sensing  of targets (also referred to as passive objects) involves target detection and, if targets are deemed to be present, estimation of their parameters \cite{will2019human}. Passive sensing include sensing of targets that do not have communication capabilities nor will aid in any form to the sensing process. Employing communication infrastructure for passive sensing of objects can enable several new use cases, such as optimizing energy consumption by controlling the \ac{IoT} devices, intruder detection, tracking of equipment among others \cite{zhang2021enabling}.  In these systems, sensing can piggyback on ubiquitous communication infra-structure there by reducing the cost for realizing these use cases. Sensing using communication signals can also ensure privacy and security  aspects compared to the existing methods which typically employ cameras to sense passive targets in indoor environments \cite{de2021convergent}. 

Methods for sensing passive objects from  the reflected signal using radars along with other onboard sensors are commonly employed in automotive use cases \cite{perez-2018-singl-frame}. These methods cannot be directly extended towards passive sensing using communication infrastructure since the sensors needed are typically not available and to mimic a traditional radar using these systems require full duplex operation to harness the reflected signals from the environment \cite{behravan-2022-introd}.  In \cite{al-qaness-2019-devic-free, Luo-2020, ma-2020-wifi-sensin, yousefi-2017-survey-behav} authors propose methods which use wireless signals for passive sensing. These methods extract features like \ac{RSSI}, \ac{CSI} or micro-Doppler shifts  from communication signal for passive sensing, based on 
mid-band ($2-10~\mb{GHz}$) carriers. 
High frequency \ac{6G} channels exhibit clustered multi-paths with each cluster pertaining to a highly reflective surface in the environment.  These channels are generally represented through environment specific ray-tracing channel models. To ensure that the conclusions drawn form the work is applicable  to many environments, stochastic geometric models, 
such as the \ac{SV} channel model \cite{saleh-1987-statis-model, li-2019-clust-based} is more appropriate. To the best of our knowledge, this model has not been adopted towards indoor passive sensing.
The passive target localization problem is also treated in the literature under the umbrella of device free localization, where the focus is only on localization and not on target detection \cite{hong-2022-learn-based, zhou-2018-devic-free}.  Typically, these works use non cellular channel models and the proposed \ac{AI} methods does not exploit the correlation in anglular domains from multiple links.  In parallel, there have been works on using \ac{RTI} for position estimation\cite{wilson-2010-radio-tomog}. In these methods, a high-resolution attenuation image caused by the presence of the object is exploited by an image estimator to arrive at the position. These methods require many communication links to get high resolution attenuation image for accurate position estimation and is not suitable for practical indoor cellular deployment.


In this paper, we develop methods that exploit the 6G infrastructure capability towards sensing of passive targets. 
The main contributions of this paper are summarized as follows. (i) An AI method that exploits the \ac{MIMO} \ac{CSI} from multiple links between transmitter and receiver towards target sensing  by perturbations in the geometric channel model. The method naturally
exploits the angular dimension of the \ac{CSI} using the rich beamforming capability of the large \ac{MIMO} array towards target sensing and parameter estimation.
(ii) Analysis of the resolution (i.e., size of the target that can be sensed),  coverage (i.e., probability of detection of a fixed size target at different spatial locations),  and position estimation accuracy, using practical indoor cellular deployments.
(ii) Comparison of the proposed position estimation method with an angle-based method to demonstrate the utility of the proposed AI-based solution.

\section{System Model}
\label{sec:sm}
In the following, we describe the system model for target sensing in the indoor environment. We assume that the deployment has  multiple links between transmit and  receive devices  having beamforming capabilities. We consider a single transmit device creating  links towards $L$ receive devices. In a typical indoor deployment the transmit devices could be a fixed anchor UE with an omni-directional antenna and the receive devices could be a \ac{BS} with beamforming capability. In the rest of the paper, we use the term transmitter and receiver to keep the discussion more general.


\subsection{Channel Model}
\label{ss:cm}
Channels in 6G systems operating at high frequency bands ($>24~\mb{GHz}$)  are sparse. Propagation paths in these channels are primarily due to the highly reflective scatterers in the environment and they arrive as clusters. Generally, deterministic channel models based on ray-tracing are commonly employed at these frequency bands. However, such channels are environment specific and does not generalize well to other environments. To overcome this and to ensure that the inference drawn from the work to be widely applicable, we adopt a stochastic geometric channel model called \ac{SV} channel model \cite{saleh-1987-statis-model, li-2019-clust-based}.  In this model, each cluster is comprised of the combination of  discrete set of rays. We consider transmissions from a signal low cost transmitter with an  omni-directional  antenna pattern and each $L$ receivers having an \ac{ULA} with $N_{r}$ elements separated by half wavelength. {Moreover, we consider a communication-centric  \ac{ISAC} system, where only a small portion of the \ac{6G} bandwidth will be used for sensing, resulting in a narrowband  channel with only spatial resolution \cite{behravan-2022-introd}.}
\subsubsection{Default Channel without Target}
During default or null state, i.e., when the object is absent, we have
\begin{equation}
  \label{eq:sv1}
  \mbf{h}^{\MS{null}}_l=\displaystyle\sum_{v=1}^{N_{\MS{cl}}}\sum_{u=1}^{N_{\MS{rays}}}\beta_{l,u,v}\mbf{\SUB{a}{rx}}(\phi_{l,u,v})G(\psi_{l,u,v}),
\end{equation}
where $\mbf{h}_l^{\MS{null}}\in\mathbb{C}^{N_{\MS{r}}}$,   $l\in\{0,\ldots,L\}$, denotes the \ac{CSI} for the link between the transmit device and $l$-th receive device in the indoor environment.  $N_{\MS{cl}}$ is the number of clusters and $N_{\MS{rays}}$ indicates the number of rays within each cluster.  The $u$-th ray of the $v$-th cluster corresponding to the $l$-th link has a complex gain $\beta_{l,u,v}$. Each ray has an angle of departure from the transmit array $\psi_{l,u,v}$ and angle of arrival at the receive array $\phi_{l,u,v}$. The transmit gain pattern is denoted by $G(\psi_{l,u,v})$, while
the receive array response is given by $[ \mbf{\SUB{a}{rx}} (\theta)]_k = e^{j\pi  k \sin(\theta)}, k\in \left[0,\ldots,N_{\MS{r}}-1\right]$. All angles are measured in the local coordinate frame of the transmitter or receivers.

\begin{figure*}
  \centering
        {\includegraphics[width = 0.9\textwidth]{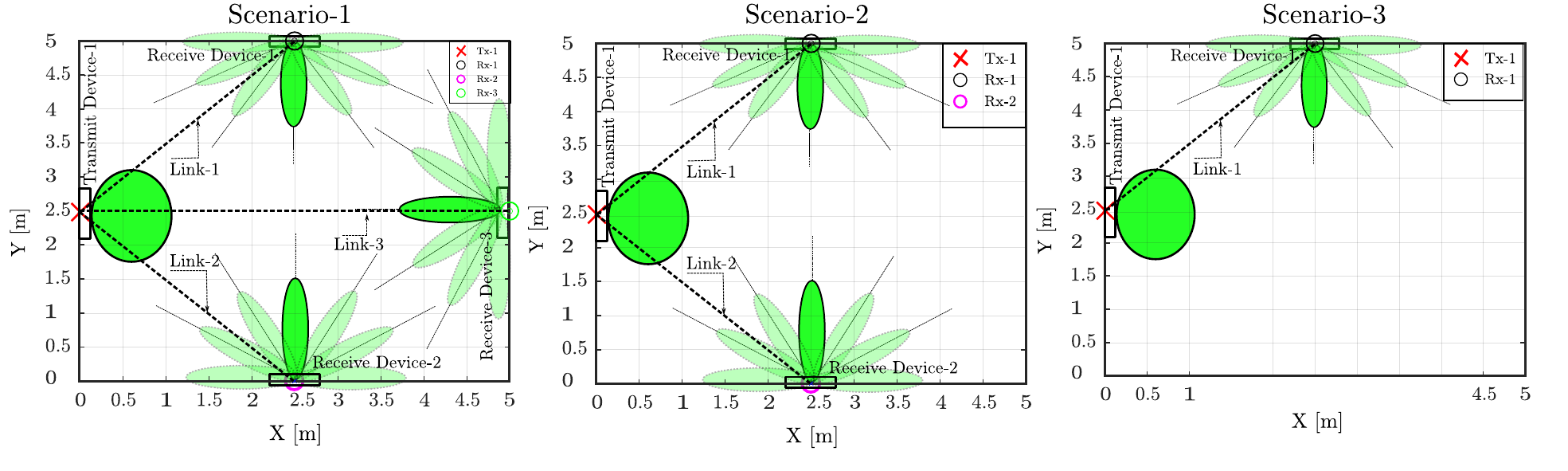}}
        \caption{Three different deployment scenarios with varied number of links, $L\in \{1,2,3\}$, are considered.}
        \label{fig:deployment} \vspace{-5mm}
    \end{figure*}

\subsubsection{Perturbed Channel with Target}
\ac{CSI} pertaining to each link  gets perturbed  uniquely when the object is placed in the environment. As shown in Fig.~\ref{fig:occlusion}, the occlusion angles $\mathcal{O}_{\MS{tx}}^1$ and $\mathcal{O}_{\MS{rx}}^{l}$ are created based on the position of the target, transmitter and receiver. Due to the high frequency of operation, we assume that the target completely blocks the rays and there is no diffraction of rays. This creates a $L+1$ convex shadow regions, namely $\mathcal{S}_{\MS{tx}}\subset \mathbb{R}^2$ behind the object as seen from the transmitter and $\mathcal{S}_{\MS{rx},l}\subset \mathbb{R}^2$  behind the object as seen from receiver $l$. 
Then,  during alternate hypothesis,  the \ac{CSI} of the channel is given by
\begin{align}
  \label{eq:sv-alt}
  \mbf{h}^{\MS{alt}}_l&=\displaystyle\sum_{u,v}\beta'_{l,u,v}\mbf{\SUB{a}{rx}}(\phi_{l,u,v})G(\psi_{l,u,v}) + \sum_{s=1}^{N_{s}} \alpha_s\mbf{\SUB{a}{rx}}(\phi_{s})G(\psi_{T}), 
\end{align}
where
\begin{equation}
  \label{eq:sv-beta}
\beta'_{l,u,v} = 
\begin{cases}  0 & \mathbf{x}(\phi_{l,u,v},\psi_{l,u,v}) \in  \mathcal{S}_{\MS{tx}} \cup \mathcal{S}_{\MS{rx},l}\\
\beta_{l,u,v}  & \mb{else},
\end{cases} 
\end{equation}
where $\mathbf{x}(\phi_{l,u,v},\psi_{l,u,v})\in \mathbb{R}^2$ is the unique location induced by the angle of departure $\psi_{l,u,v}$ from the transmitter and angle of arrival $\phi_{l,u,v}$ from the $l$-th receiver. {The second term of \eqref{eq:sv-alt} represents the contribution due to the scattering from the target resulting in $N_s$ rays arriving at the receiver, having complex  gains  $\alpha_s$, angles of arrival $\phi_s$ and a fixed angle of departure $\psi_{T}$. Here,  $\psi_{T}$ denotes the angle of the impinging ray from the transmitter to the center of the target.} 

So far we assumed a single target of interest in the scene during alternate hypothesis. However when there are multiple targets ($T>1$), the perturbed \ac{CSI} is  due to the creation of $T(L+1)$ shadow regions, together with the new reflection paths reaching the receivers due to the scattering from $T$ targets. Without loss of generality, the above  proposed methods can be extended to the multi-target scenarios with much richer interaction between the objects and the impinging rays.

\subsection{Deployment Model}
\label{ss:dp}
We consider an indoor deployment in a $25~\mb{m}^2$ area with a transmit device (fixed anchor UE) having an  omni-directional antenna ($N_t=1$) and multiple receive devices (BSs) having an \ac{ULA} with $N_{\MS{r}}=8$ antennas.  We place the transmit and receive device such that the boresight direction is normal to the walls as shown in Fig.~\ref{fig:occlusion}. Each receiver has beamforming capability to scan between $-\pi/2 \mb{ to } +\pi/2$ using $N_b$  beams. An illustration of three deployment scenarios with number of links, $L\in \{1,2,3\}$ with receivers performing a beam scan using $N_b=7$ beams is shown in the Fig.~\ref{fig:deployment}. During each \ac{CPI}, \ac{CSI} is captured in all the $N_b=7$ angular dimensions synchronously for each link and transferred to an AI agent where detection and parameter estimation on the passive target is performed.

\section{Methods}
\label{sec:methods}
The complex relationship between high dimensional \ac{CSI} space to the target detection and parameter estimation can be learned by AI methods directly from data without modeling. In this section, we discuss the AI methods and required data pre-processing for the sensing problem. 
\subsection{Data  Preprocessing and AI Architecture }
\label{ss:ai}
    \begin{figure*}[t]
      \centering
      {\includegraphics[width = 0.7\textwidth]{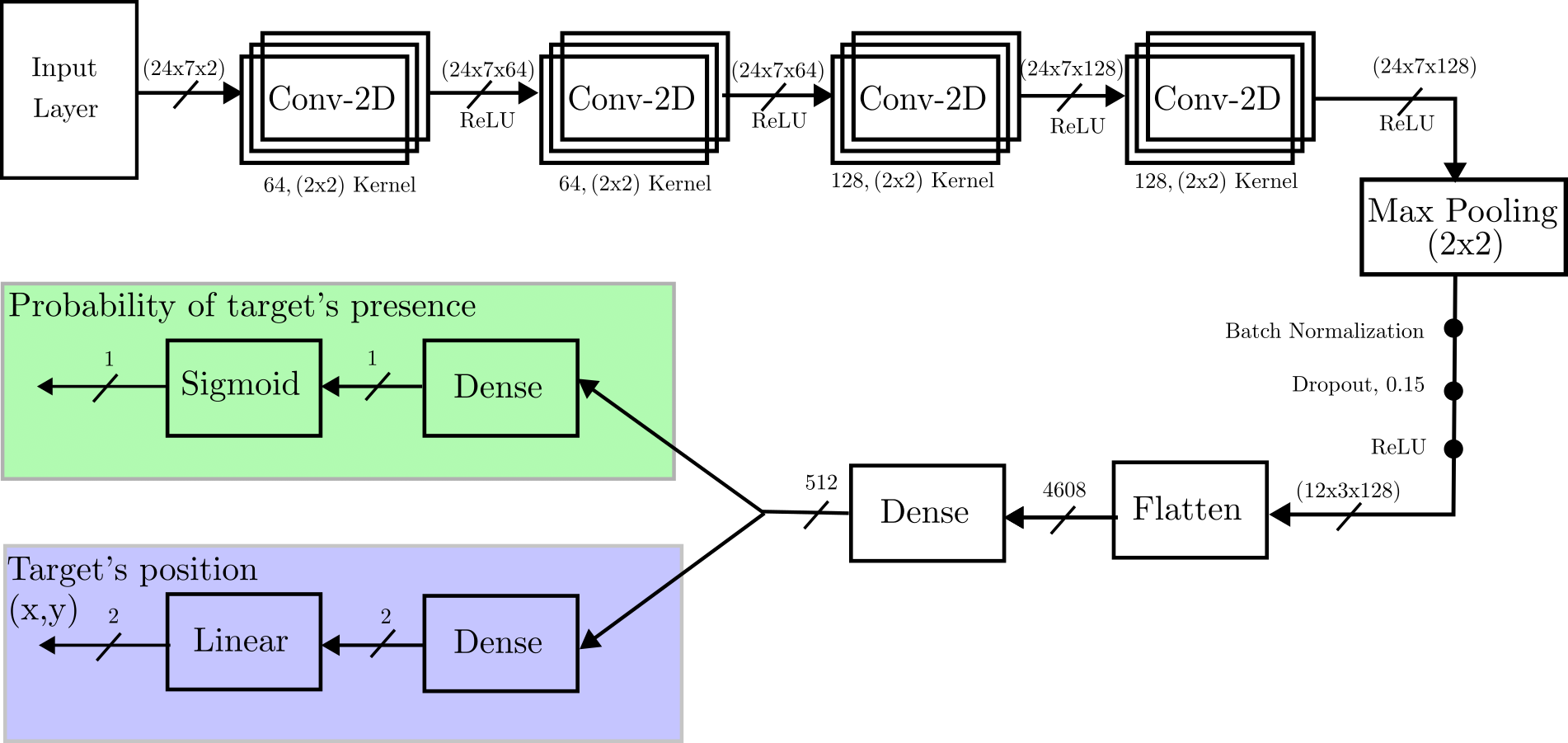}}
      \caption{AI pipeline for target detection and position estimation considering  scenario-3 ($L= 3$). Input to the model is a 2D-CSI frame, $\mbf{H} \in \mathbb{C}^{L N_{\MS{r}} \times N_b}$. With $L=3,N_r=8,N_b=7$, and separating real and imaginary values into different channels, we have the 2D-CSI frame dimension of $(24\times7\times2)$. Both target detection and position estimation share the same network except at the last two layers.}\vspace{-5mm}
      \label{fig:merged_model}
    \end{figure*}
We represent the \ac{CSI} for each \ac{CPI} in the form of a 2D frame. which is fed to an \ac{AI} based imaging processing pipeline consisting of stacked \ac{CNN} to extract relevant features. Similar to \ac{AI} based image processing, the pipeline is supervised to learn the relation between input 2D-CSI space  to output space. The structure of the CSI data  is used to tune the hyper parameters of the AI - pipeline. Tuning is done in such a way to have the network as shallow as possible at the same time yields good performance so that it can be used on an embedded platforms. We call this tuned CNN network as CsiSenseNet, and is shown in Fig.~\ref{fig:merged_model}. Both target detection and position estimation pipelines share the same network except for the last two layers shown in green shaded area for target detection and blue shaded area for position estimation.

The CSI for all the $L$  links are concatenated in the horizontal dimension, that is for a given receiver beamforming angle, $\theta_i$ at all receivers,
\begin{equation}
  \label{eq:dp-1}
\mbf{h}_{\theta_i} = [ \mbf{h}_{1,\theta_i}|\mbf{h}_{2,\theta_i}|\ldots|\mbf{h}_{L,\theta_i}]^{\MS{T}}\in \mathbb{C}^{L \cdot N_{\MS{r}} \times 1},
\end{equation}
where $|$ is the concatenation operation,  $\mbf{h}_{\theta_i}$ is the aggregated CSI in a particular angular direction $\theta_i$ and  $\mbf{h}^{\MS{T}}_{l,\theta_i} \in \mathbb{C}^{N_{\MS{r}}}, l\in \{1,\ldots,L\}$ denotes the \ac{CSI} for $l$-th link. The collected CSI from different angular directions are further concatenated in the vertical dimension forming a 2D-CSI frame

\begin{equation}
  \label{eq:dp-3}
  \mbf{H} = [  \mbf{h}_{\theta_1}^{\MS{T}}|  \mbf{h}_{\theta_2}^{\MS{T}}|\ldots |\mbf{h}_{\theta_{N_b}}^{\MS{T}}  ]^{\MS{T}} \in \mathbb{C}^{L N_{\MS{r}} \times N_b}.
  \end{equation}
Both pipelines are separately trained for target detection and position estimation respectively.
\subsection{Target Detection}
\label{ss:td}
For target detection, several realizations of channel  $\mbf{H}$ are generated using a simulator for both hypotheses (i.e., with and without a target). A labeled training set consisting of $M$ records $(\mbf{H}_i,\mb{hyp}_i)\suchthat i=1,2,\ldots,M$ with $\mbf{H}_i$ as channel realization and $\mb{hyp}_i$ as hypothesis is used 
to supervise the target detection (green shaded) part of the AI pipeline shown in Fig.~\ref{fig:merged_model}. Detection network is trained to  minimize binary cross-entropy loss.

\subsection{Position Estimation}
\label{ss:pe}
The position estimation part of the CsiSenseNet shown in the blue shaded area of Fig~\ref{fig:merged_model} has two neuron output (for $X$ and $Y$ coordinate estimates) with a linear activation. Similar to target detection a labeled data set consisting of  $(\mbf{H}_i,\mbf{p}_i)\suchthat i=1,2,\ldots,M$ with $\mbf{p}_i\in \mathbb{R}^2$ representing the position of the target for channel realization $\mbf{H}_i$, is used to supervise the position estimation network.

We use angle-based position estimation to compare performances with the proposed CsiSenseNet based position estimator. Since in the representative deployment scenarios shown in Fig.~\ref{fig:deployment}, the receivers employ multiple antennas and are beamforming capable, the baseline method identifies the angular direction of the beam which is observing maximum perturbation (attenuation) from multiple receivers for triangulating to the position.
    \begin{figure}
      \centering
      {\includegraphics[width = 0.4\textwidth]{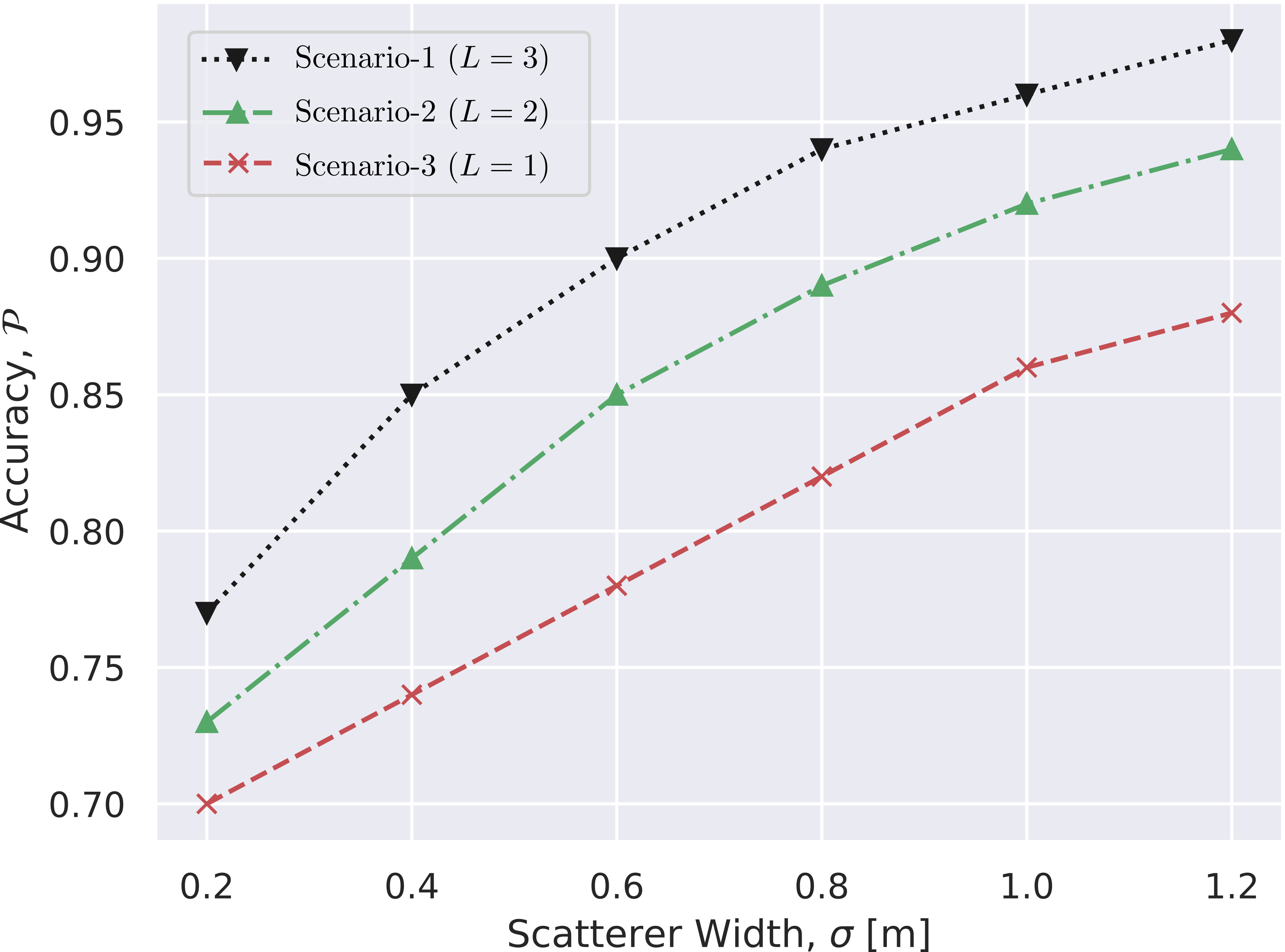}}
      \caption{Accuracy versus target size for different deployment scenarios of Fig~\ref{fig:deployment}. }\vspace{-4mm}
      \label{fig:resolution}
    \end{figure}

    \begin{figure*}
  \centering
  {\includegraphics[width = 0.9\textwidth]{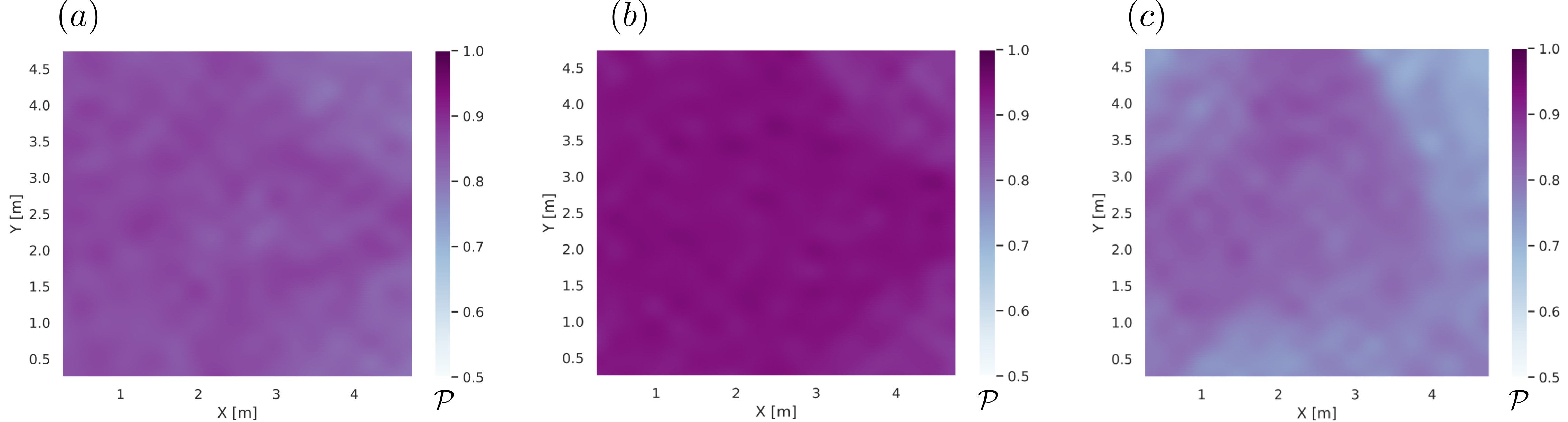}}
  \caption{Coverage of the proposed AI detector. (a) Coverage for deployment scenario-1 ($L=3$) with target size $\sigma=0.5~\mb{m}$ (b) Coverage for deployment scenario-1 ($L=3$) with target size $\sigma=0.8~\mb{m}$ (c) Coverage for deployment scenario-3 ($L=1$) with target size $\sigma=0.8~\mb{m}$.  }\vspace{-5mm}
  \label{fig:coverage}
\end{figure*}

    
\section{Simulation Results}
\label{sec:results}

\subsection{Simulation Setting}
\label{sec:ss}
We use \ac{MFM} simulator discussed in  \cite{mfm}  to create the deployments shown in the Fig~\ref{fig:deployment}. {Due to the high absorption characteristics of high frequency 6G channels, we modify the \ac{SV} channel model to have single bounce reflection from scatter to the receiver, as detailed in the Appendix.} We consider beamforming only in the azimuth direction and assume circular shapes for the target to aid in analysis.  The proposed methods can be easily extended to have beamforming in both azimuth and elevation with arbitrary shaped targets. Although we consider a single target in the simulations, the \ac{AI} pipeline can be trained with data from a much larger input domain space having many targets of interest at various positions for multi-target sensing. We configure the simulator as shown in the Table \ref{tab:params}. 

In terms of performance metrics, we first of all consider the accuracy score $\mathcal{P}$: 
\begin{align}
    \mathcal{P} & = 1-(p(\mathrm{target} | \mathrm{null})p(\mathrm{null})+ p(\mathrm{null} | \mathrm{target})p(\mathrm{target})),
\end{align}
which is estimated empirically during testing. Using $\mathcal{P}$ we define resolution as the size of the target that can be sensed with an accuracy score higher than a threshold (i.e.,$\mathcal{P}>\gamma$) and coverage as variation of $\mathcal{P}$ at different spatial points for a fixed size target. Secondly, we consider the \ac{CDF} of the positioning error, i.e., $F_{\mathcal{E}}(\varepsilon)$, where $\varepsilon=\norm{\hat{\mbf{p}}-\mbf{p}}$, in which $\norm{\cdot}$ denotes the L2 norm and  $\hat{\mbf{p}} \in \mathbb{R}^2$ is the position estimate of the true position, $\mbf{p}$.
\subsection{Results and Discussion}
We now proceed to evaluate the impact of the size of the target and the spatial coverage for different numbers of receivers. Then we evaluate the target positioning performance and compare to a model-based baseline. 

\subsubsection{Resolution Analysis}
\label{sec:ra}
We analyzed the size of the target required to create sufficient CSI perturbation to be detected by the AI agent. First, we generate $2000$ CSI realizations for each hypothesis and size by placing object at 1000 random positions within  a $25~\mb{m\textsuperscript{2}}$ indoor area.  A $70/30$ split is done to  train and validate the target detection part of the CsiSenseNet AI pipeline. 
Then  we drop objects with varying size having diameter, $\sigma$ from $0.2$ to $1.2$ at  $700$ random positions drawn from a $25~\mb{m\textsuperscript{2}}$ area to assess the accuracy of the AI prediction. The accuracy score, $\mathcal{P}$, of the AI detector for the representative deployment scenarios in Fig.~\ref{fig:deployment} is shown in the Fig.~\ref{fig:resolution}.   The performance of the detector improves with $L$ and for a given deployment, larger sized targets can be sensed with higher accuracy. For a passive object such as human, who  has an approximate width of about $0.8~\mb{m}$ can be detected with more than $90$ percent accuracy with $L>2$. 
\begin{table}
  \centering
  \begin{tabular}{|c|c|}
\hline
    \textbf{Simulation Parameter}&\textbf{Value}  \\
\hline
    Number of links & $L=1,2,3$\\
    Target size (diameter) & $\sigma=0.5,0.8,1.0,1.5$ \\
    Tx antenna array & $1\times 1$\\
    Rx antenna array & $1\times 8$\\
    Number of beams & $N_b=7$ \\
    Beam sweep angles & $\{ -\nicefrac{\pi}{2},-\nicefrac{\pi}{3},-\nicefrac{\pi}{6},$ $0,  \nicefrac{\pi}{6},\nicefrac{\pi}{3}, \nicefrac{\pi}{2}\}$\\
    Channel model & Modified \ac{SV} model,  $N_{\MS{cl}}, N_{\MS{rays}}, N_s=3, 5,1$ \\
\hline                      
  \end{tabular}
  \caption{Simulation parameters}
  \label{tab:params}\vspace{-5mm}
\end{table}

\begin{figure*}[t]
  \centering
  {\includegraphics[width=1\textwidth]{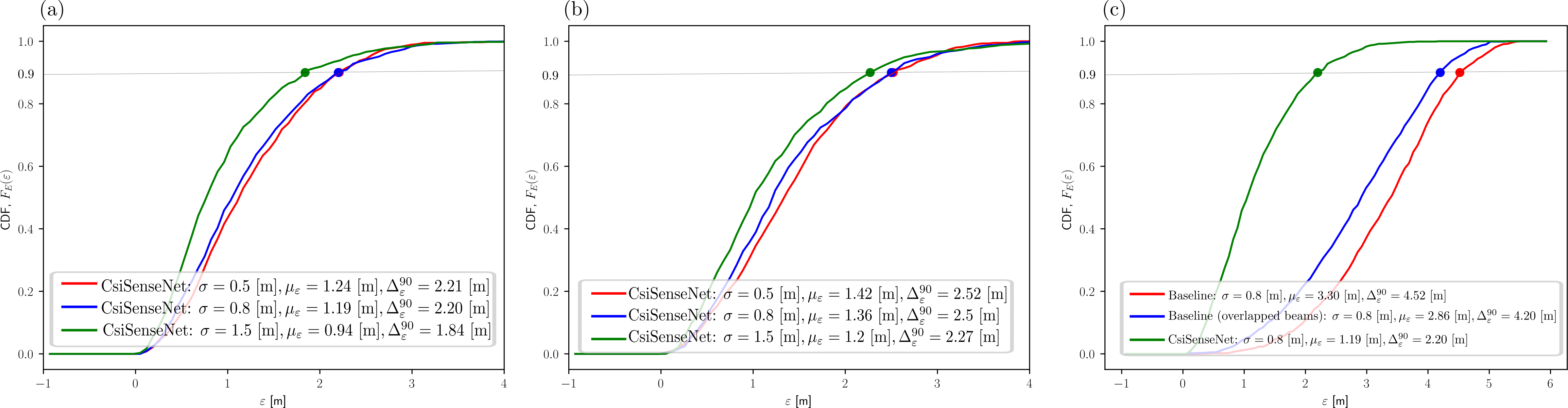}}
  \caption{Positioning performance: (a) and (b) shows the performance of CsiSenseNet in terms of mean ($\mu_{\varepsilon}$), 90-percentile ($\Delta^{\MS{90}}_{\varepsilon}$) and \ac{CDF} ($F_{\mathcal{E}}(\varepsilon)$) of the position error, $\varepsilon$ for representative deployments, scenarios-1 and  scenario-3 respectively for different target sizes. (c) Comparison of the positioning performance between baseline and CsiSenseNet methods for target size, $\sigma=0.8$. }\vspace{-5mm}
  \label{fig:PositioningConsolidated}
\end{figure*}

\subsubsection{Coverage Analysis}
\label{sec:ca}
The separation of the distribution of CSI matrix under null hypothesis (without targets) and alternate hypothesis (with target) depends on the position of the target. Positions closer to the transmitter or receiver node creates more CSI perturbation in alternate hypothesis than the targets which are farther. For example, objects in the direction near the endfire of an array are less likely to be detected than objects near the broadside. 
Therefore, we can define coverage of the  sensing method for a given sized target in terms of probability of target detection at various positions.
To assess the coverage, we train the target detection part of the CsiSenseNet by generating 2000 CSI realizations for both hypotheses by placing the fixed size object at the center of various quantized bins of $0.0625~\mb{m\textsuperscript{2}}$ of a $25~\mb{m\textsuperscript{2}}$ indoor area. The performance of such a trained agent is evaluated using $700$ new CSI realization for both hypothesis at each of the quantized bins of $0.0625~\mb{m\textsuperscript{2}}$ from total indoor area of $25~\mb{m\textsuperscript{2}}$ for representative deployment scenarios in Fig.~\ref{fig:deployment}. The coverage of the proposed sensing method is shown in  Fig.~\ref{fig:coverage} for representative deployment scenarios. The coverage is good at positions closer to the transmit and receive antennas, and also along the beam directions. Also comparing Fig~\ref{fig:coverage}(a) and Fig.~\ref{fig:coverage}(b) notice that the coverage depends on the size with larger sized target having better coverage.

\subsubsection{Position Estimation with CsiSenseNet}
\label{sec:pe}
In this section, we present the results for the proposed position estimation of the target using  \ac{CSI}  gathered from multiple links and compare its performance with baseline method.
%
For a fixed target-size, $2000$ CSI realizations at each quantized bin positions of resolution $0.0625~\mb{m\textsuperscript{2}}$ is captured similar to Section \ref{sec:ca}, which is then used to train the position estimation part of the CsiSenseNet. We then drop the objects of various size ($\sigma$) at  $1000$ random positions drawn from a $25~\mb{m\textsuperscript{2}}$ area to access the accuracy of the position estimation. The Fig.~\ref{fig:PositioningConsolidated}(a) and Fig.~\ref{fig:PositioningConsolidated}(b) shows the performance in terms of mean position error $\mu_{\varepsilon}$, $90\mb{-percentile error } \Delta^{\MS{90}}_{\varepsilon}$, and \ac{CDF} of position-error $F_{\mathcal{E}}(\varepsilon)$ for different deployment scenarios and target sizes. From Fig.~\ref{fig:PositioningConsolidated}(a) and Fig.~\ref{fig:PositioningConsolidated}(b), larger target size and more number of links in the deployment reduces the position uncertainty.
\subsubsection{Position Estimation with Baseline Method}
\label{sec:baseline}
 The performance of the baseline method described in Section~\ref{ss:pe} is as shown in  Fig.~\ref{fig:PositioningConsolidated}(c). The red plot in Fig.~\ref{fig:PositioningConsolidated}(c) is the performance of the baseline algorithm using  $7$ non-overlapping beams to scan the space  $(-\pi/2,+\pi/2)$ as shown in Fig.~\ref{fig:deployment}. The high position uncertainty in this method is  due to:
\begin{enumerate}[label=(\alph*)]
\item The representative deployment scenarios use  $N_{\MS{r}}=8$ antennas at receiver, which yields approximate angular resolution of $30$ degrees which is rather high and creates greater uncertainty while triangulating the angles towards position
\item The beams are not over-lapping which creates the large spatial regions without coverage
\item Due to the geometry of receiver placements and the target position, it could block multiple adjacent beams leading to angular uncertainty and  inferior position estimates.
\end{enumerate}
To address the issue described in (b) above, we created  overlapped beams with beam width $30$ degrees with stride of one degree to span   $(-\pi/2,+\pi/2)$ resulting in $180$ overlapped beams. The performance of the angle based estimator with this modification is shown in the blue plot of Fig.~\ref{fig:PositioningConsolidated}(c). This modification to the baseline algorithm reduced the average position error, $\mu_{\varepsilon}$ from $3.30~\mb{[m]}$ to $2.86~\mb{[m]}$. The CsiSenseNet outperforms the angle based methods because the AI agent learns the spatial correlation between the perturbance in a higher dimension \ac{CSI} space for each angular dimension across multiple receivers towards position estimation. 

\section{Conclusions}
\label{sec:conclusion}

Passive sensing of targets using ubiquitous communication infrastructure provides several beneﬁts without compromising on privacy and security as in the camera aided sensing systems. This paper describes a multistatic indoor sensing system which exploits perturbation patterns from inserted objects in the CSI of multiple links towards detection and position estimation. A shallow CNN based AI network called CsiSenseNet is developed to exploit these patterns towards target sensing. Results show that larger objects are easier to detect with higher accuracies. The performance of the proposed method to estimate the sensed target's position improves with the objects size and outperforms angle based methods. Objects inserted close to transmitter or receiver or along scanned beam directions are easier detected than the objects in other places. Increasing the number of links improves detection and position accuracy. Based on the results, the proposed methods can be used for sensing humans sized objects with good accuracy using indoor cellular deployment.

\section*{Appendix: Modified \ac{SV} Model}
\begin{figure}[h!]
      \centering
      {\includegraphics[width = 0.6\columnwidth]{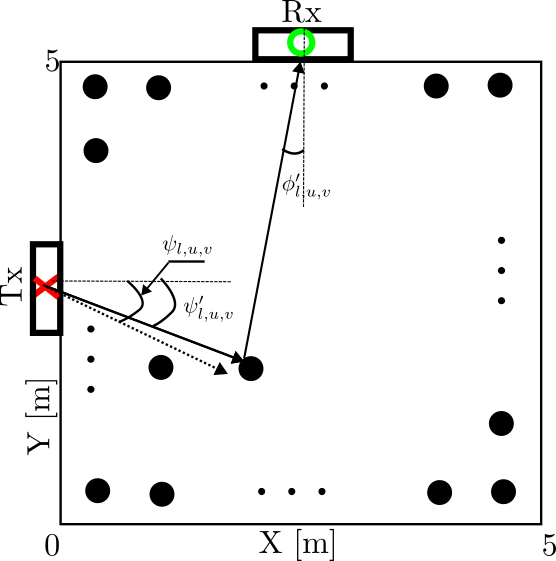}}
      \caption{Modified \ac{SV} model with single bounce reflections. }
      \label{fig:modified_SV}\vspace{-5mm}
    \end{figure}

In order to modify the \ac{SV} to only have single bounce reflections, we discretize the possible angle of departures into a set of angles pointing to a fine grid of points with $0.0625~\mb{m\textsuperscript{2}}$ resolution. The stochastically generated angle of departure ($\psi_{l,u,v}$)  from the \ac{SV} model are quantized to closest discretized angle ($\psi'_{l,u,v}$) corresponding to the quantized grid point as shown in Fig.~\ref{fig:modified_SV}. By using the  location of the receiver and the grid point, the angle of arrival ($\phi'_{l,u,v}$) to the receiver is computed from geometry. 
\section*{Acknowledgments}
This work has been partly funded by the European Commission through the H2020 project Hexa-X (Grant Agreement no. 101015956). The authors gratefully acknowledge  feedback and advise from Robert Baldemair.

\balance 
\bibliography{main}
\bibliographystyle{IEEETran}
 
\end{document}